\documentclass[12pt,sort&compress]{article}
\usepackage{ae,aecompl}
\usepackage[T1]{fontenc}
\usepackage[latin9]{inputenc}
\usepackage[active]{srcltx}
\usepackage{color}
\usepackage{array}
\usepackage{float}
\usepackage{booktabs}
\usepackage{multirow}
\usepackage{amsmath}
\usepackage{graphicx}
\usepackage[numbers]{natbib}

\makeatletter

\providecommand{\tabularnewline}{\\}

\usepackage{aecompl}\usepackage{epstopdf}

\usepackage[labelfont=bf]{caption} 
\usepackage{caption} 

\makeatother

\begin{document}
\title{\textbf{\textcolor{black}{\Large{}Point-defect avalanches mediate
grain boundary diffusion}}\textbf{\textcolor{black}{{} }}\textcolor{black}{\date{}}}
\maketitle
\noindent \begin{center}
\textcolor{black}{Ian Chesser$^{1}$ and Yuri Mishin$^{1}$}
\par\end{center}

\section*{\textcolor{black}{Abstract}}

Grain boundary (GB) diffusion in polycrystalline materials is a physical
phenomenon of great fundamental interest and practical significance.
Although the accelerated (``short circuit'') atomic transport along
GBs has been known for decades, the current atomic-level understanding
of GB diffusion remains poor. Experiments can measure numerical values
of GB diffusion coefficients but provide little information about
the underlying mechanisms. Previous atomistic simulations focused
on relatively low temperatures when the GB structure is ordered or
relatively high temperatures when it is highly disordered. Here, we
report on molecular dynamics simulations of GB diffusion at intermediate
temperatures, which are most relevant to applications. One of the
surprising results of this work is the observation of strongly intermittent
GB diffusion behavior and its system-size dependence unseen in previous
work. We demonstrate that both effects originate from an intermittent,
avalanche-type generation of point defects causing spontaneous bursts
of GB diffusivity mediated by highly cooperative atomic rearrangements.
We identify the length and time scales of the avalanches and link
their formation to the dynamic heterogeneity phenomenon in partially
disordered systems. Our findings have significant implications for
future computer modeling of GB diffusion and mass transport in nano-scale
materials.

\textcolor{black}{\vfill{}
}

\noindent \textcolor{black}{$^{1}${Department of Physics and Astronomy,
MSN 3F3, George Mason University, Fairfax, Virginia 22030, USA}.
Correspondence and requests for materials should be addressed to I.C.
(ichesser@gmu.edu) or Y.M. (email: ymishin@gmu.edu).}

\textcolor{black}{\newpage}

\section*{\textcolor{black}{Introduction}}

\noindent \textcolor{black}{Most technological materials are composed
of differently oriented crystallites (grains) separated by a network
of grain boundaries (GBs) \citep{Balluffi95}. Atomic transport along
GBs is known to be much faster than in the perfect lattice }\citep{Kaur95}\textcolor{black}{.
This }``short-circuit'' GB diffusion controls many processes in structural
and functional materials, including microstructure development, phase
transformations, and high-temperature modes of plastic deformation
and fracture \citep{Kaur95,Balluffi95}. In addition to the practical
significance, GB diffusion is a phenomenon of great fundamental interest.
Being in a frustrated state squeezed between misoriented crystals,
the quasi-two-dimensional GB structure represents a unique state of
matter ``transitional'' between ordered and disordered. Due to the
structural complexity, the current understanding of GB diffusion mechanisms
remains very limited. Experiments can only measure numerical values
of GB diffusion coefficients but provide little information about
the microscopic mechanisms. Hence, practically all existing knowledge
about GB diffusion mechanisms comes from atomistic computer simulations
\citep{mdgbreview}.

Previous simulations have only addressed two limiting cases of relatively
low and relatively high temperatures. It is now established that at
low temperatures ($T\ll T_{m}$, where $T_{m}$ is the melting point),
GBs support both vacancies and interstitials, which can exist in a
variety of structural forms, have strongly site-dependent but on average
lower formation energies than in the lattice, and can contribute equally
to GB diffusion \citep{sorensen2000diffusion,suzuki2003atomistic,suzuki2005atomic,mishin2015atomistic}.
GB atoms can diffuse by many different but predominantly collective
mechanisms, typically in the form of a chain of highly coordinated
displacements that can be open-ended (string) or looped (ring). In
the other limit, when temperature approaches the bulk melting point,
the GB structure becomes highly disordered. Some GBs premelt by transforming
into a liquid-like layer near $T_{m}$ \citep{suzuki2005atomic,mishin2015atomistic,gbmdhpnas}.
The notion of individual point defects breaks down, but the diffusion
mechanisms remain collective and closely resemble those in disordered
bulk systems such as supercooled liquids, glasses, and crystalline
super-ionic materials \citep{donati1998stringlike,derlet2021viscosity,AlSm1,annamareddy_superionic}.
Accordingly, the statistical methods developed for the characterization
of atomic dynamics in disordered bulk systems were successfully applied
to study diffusion mechanisms in disordered GBs \citep{mishin2015atomistic,gbmdhpnas}.
Such studies revealed a remarkable similarity between the two types
of systems, including the existence of string-like diffusion mechanisms
and the dynamic heterogeneity occurring on multiple length and time
scales.

Little is known about GB diffusion at intermediate temperatures, at
which the GB structure is not liquid-like but rather retains a significant
degree of structural order. It is presently unknown if the concepts
of dynamic heterogeneity and dynamic facilitation developed for disordered
systems can be applied to such boundaries. If they can, the length
and time scales of the heterogeneous dynamics remain unexplored. It
should also be noted that the intermediate temperatures are more relevant
to the service conditions of many technological materials than the
two limiting cases mentioned above.

This work aims to fill this knowledge gap by studying the diffusion
mechanisms in metallic GBs at intermediate temperatures. A surprising
outcome of the study is the observation of a strongly intermittent
diffusion behavior and its system-size dependence unseen in previous
work. We show that both effects originate from an intermittent, avalanche-type
generation of point defects causing spontaneous bursts of GB diffusivity.
We carefully characterize the length and time scales of the avalanches
and show that they fully explain the observed intermittent diffusion
behavior and its size dependence. We discuss possible implications
of our findings for the computer modeling of GB diffusion and for
atomic transport in nano-scale systems.

The primary material chosen for this study is face-centered cubic
(FCC) Al. Four different Al GBs were selected as shown in Table \ref{tab:1}.
Two of them are the high-angle tilt $\Sigma3$ $\left\langle 110\right\rangle $
and $\Sigma17$ $\left\langle 100\right\rangle $ GBs, where $\Sigma$
is the reciprocal density of coincident sites and the angular brackets
indicate the tilt axis \textcolor{black}{\citep{Balluffi95}}. The
third boundary is a low-angle twist GB with the $\{100\}$ GB plane,
and the last one is a high-angle asymmetric incommensurate (HAAI)
boundary obtained by joining two crystal planes with incommensurate
periodicities. Both the $\Sigma 3$ and HAAI boundaries have relatively low energies and are common
in polycrystalline Al \citep{GBCD2004}. These four boundaries represent
diverse GB crystallographies, symmetries, and atomic structures. Most
of the results reported below are for the $\Sigma3$ and $\Sigma17$
GBs, with two remaining boundaries left for cross-checking. We use
classical molecular dynamics (MD) with atomic interactions described
with the interatomic potential \citep{Almendelev} referred to here
as Al-M. To demonstrate that our results are not a specific feature
of a particular interatomic potential, some of the simulations are
repeated with two other Al potentials. The results were found to be
well-consistent with those obtained with the Al-M potential. Furthermore,
additional simulations were performed for the $\Sigma3$ GB in FCC
Ni and $\Sigma17$ GB in body-centered cubic (BCC) Fe, which revealed
similar diffusion behaviors. These cross-checks confirm the robustness
of our conclusions across different metals with different crystalline
structures. 

\section*{\textcolor{black}{Results}}

\begin{figure}[H]
\noindent \begin{centering}
\includegraphics[width=0.63\textwidth]{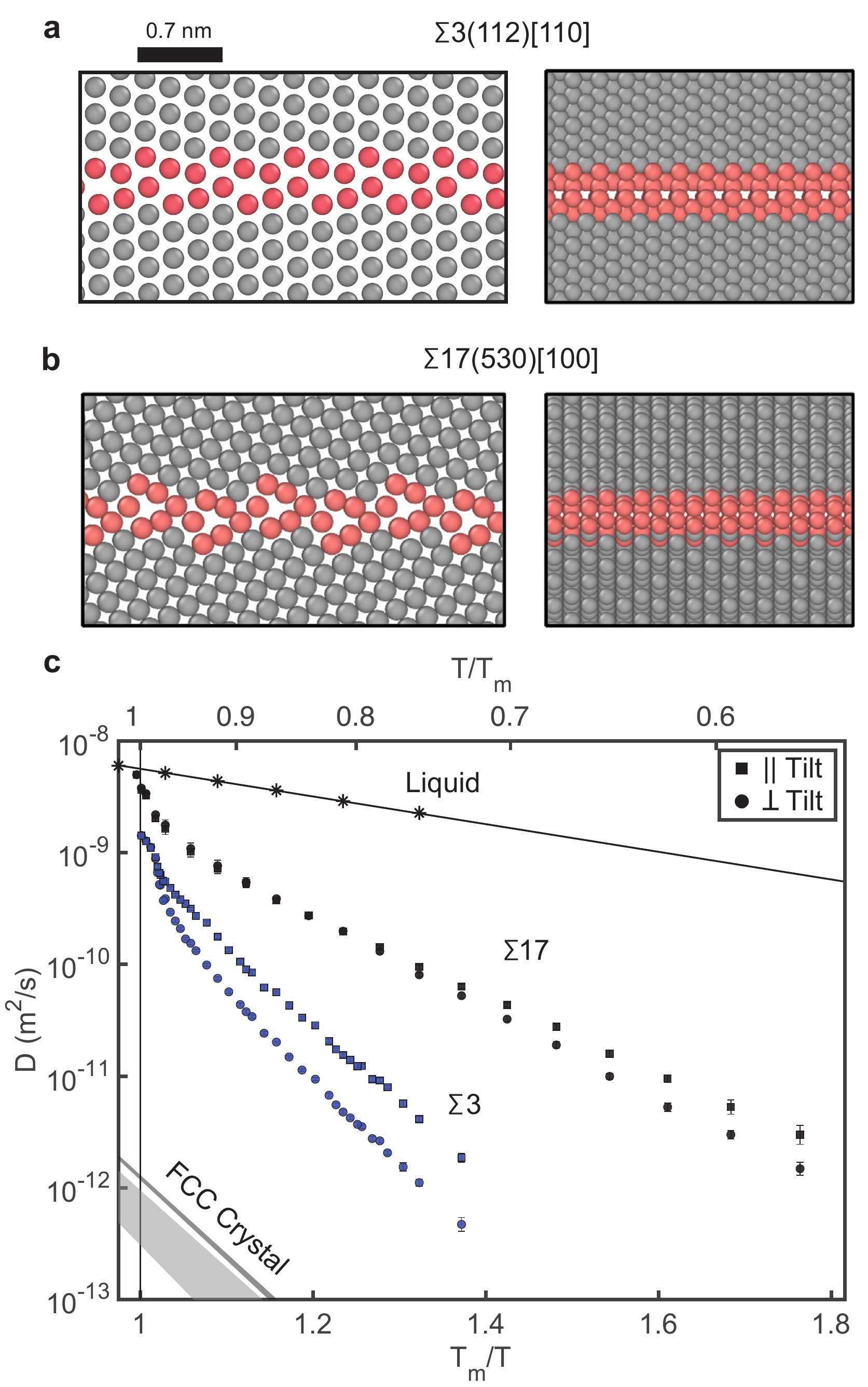}
\par\end{centering}
\noindent \centering{}\caption{Atomic structure and diffusion coefficients in the Al $\Sigma3$ and
$\Sigma17$ GBs. Panels (a) and (b) show the optimized GB structures
as viewed parallel (left column) and normal (right column) to the
tilt axis. Atoms identified with perfect FCC coordination via common
neighbor analysis are colored grey and all other atoms are colored
red. (c) Arrhenius diagram of GB diffusion coefficients parallel ($D_{||}$)
and perpendicular ($D_{\bot}$) to the tilt axis. The vertical dashed
line indicates the melting point predicted by the interatomic potential.
Shown for comparison are the computed diffusivities in liquid Al and
the experimental (dashed stripe) \citep{dais1987nuclear,Zhong:2020we}
and computed (lower stripe) \citep{Hooshmand:2020tu} literature data
for self-diffusion in FCC Al.}
\label{fig:fig1}
\end{figure}

The $\Sigma3$ and $\Sigma17$ GB structures optimized with the Al-M
potential are shown in Fig.~\ref{fig:fig1}a,b. The GB construction
and optimization protocol is explained in the Methods section. The
$\Sigma3$ GB, also known as incoherent twin boundary, is composed
of asymmetric kite-shaped structural units. This structure closely
matches the high-resolution transmission electron microscopy observations
\citep{medlin1992hrtem,medlin1993migration} and prior simulations
\citep{wang2011shear}. The asymmetry arises from a small shift of
the upper grain relative to the lower along the $\left\langle 111\right\rangle $
axis ($x$-direction in the sample frame), whose magnitude ($0.7$
angstroms) is close to the experimental value ($0.66\pm0.07$
angstroms) \citep{medlin1993migration}). This boundary has also
a symmetric structure (zero shift) \citep{medlin1993migration,marquis2004finite},
but the latter has a higher energy. We thus use the asymmetric structure
for the diffusion simulations. The $\Sigma17$ GB has a zig-zag structure
with the kite-unit spacing ($0.59$ nm) and energy consistent with
prior simulations \citep{kojuAlMg}. This boundary was previously
used to study premelting diffusion mechanisms in copper \citep{mishin2015atomistic}.

Self-diffusion coefficients in the $\Sigma3$ and $\Sigma17$ GBs
were extracted from MD trajectories using the Einstein relation as
detailed in Methods. The results are summarized in the Arrhenius diagram
in Fig.~\ref{fig:fig1}c. For comparison, the diagram also includes
the diffusion coefficients in liquid Al computed in this work and
the diffusion coefficients in FCC Al taken from the literature. In
agreement with prior work \citep{suzuki2005atomic}, the GB diffusion
coefficients are intermediate between those in supercooled liquid
and in the perfect crystal. Both GBs exhibit diffusion anisotropy
with faster diffusion along the tilt axis ($y$-axis) than normal
to the tilt axis ($x$-axis). Diffusion in the $\Sigma3$ GB displays
a larger degree of anisotropy and is over an order of magnitude slower
than diffusion in the $\Sigma17$ GB (except close to the melting
temperature when the two diffusivities tend to converge). This diffusion
behavior correlates with the lower energy of the $\Sigma3$ GB (418
mJ/m$^{2}$) relative to the $\Sigma17$ GB (488 mJ/m$^{2}$).

Below the homologous temperature of $0.9\ T_{m}$ ($T_{m}=929$ K
is the Al melting point predicted by the Al-M potential), the GB diffusion
coefficients follow the Arrhenius relation 
\begin{equation}
D_{i}=D_{0,i}\exp\left(-\dfrac{E_{i}}{k_{B}T}\right),\label{eq:1}
\end{equation}
where the index $i$ refers to the directions parallel ($||$) or
normal ($\bot$) to the tilt axis, $D_{0,i}$ is the entropic prefactor,
$E_{i}$ is the activation energy of diffusion, and $k_{B}$ is Boltzmann's
constant. Table \ref{tab:Tact} summarizes the Arrhenius parameters
for each GB obtained by fitting Eq.(\ref{eq:1}) to the data below
$0.9\ T_{m}$. For the $\Sigma3$ GB, the average activation energy
(1.38 eV) is significantly higher than for the $\Sigma17$ GB (0.77
eV) and comparable to the literature data for lattice diffusion (1.25-1.45
eV) \citep{Hooshmand:2020tu}. At high temperatures, a premelting
regime is observed in both GBs as evident from the rapid increase
in the GB diffusivity with temperature. The $\Sigma17$ GB fully premelts
and its diffusivity comes close to that in liquid Al near $T_{m}$,
which is consistent with previous work on this GB in copper \citep{suzuki2005atomic,kojuAlMg,hickman2016disjoining}.
The $\Sigma3$ GB also premelts, but its diffusivity remains below
the bulk liquid diffusivity. This work is focused on diffusion at
intermediate temperatures, $T<0.8\ T_{m}$ ($T<750$ K), at which
both GBs are relatively flat and immobile on the MD timescale. This
allows us to cleanly analyze the diffusion mechanisms without the
added complications from GB roughening, migration, or premelting.

\begin{figure}[H]
\centering\leavevmode \includegraphics[width=0.77\textwidth]{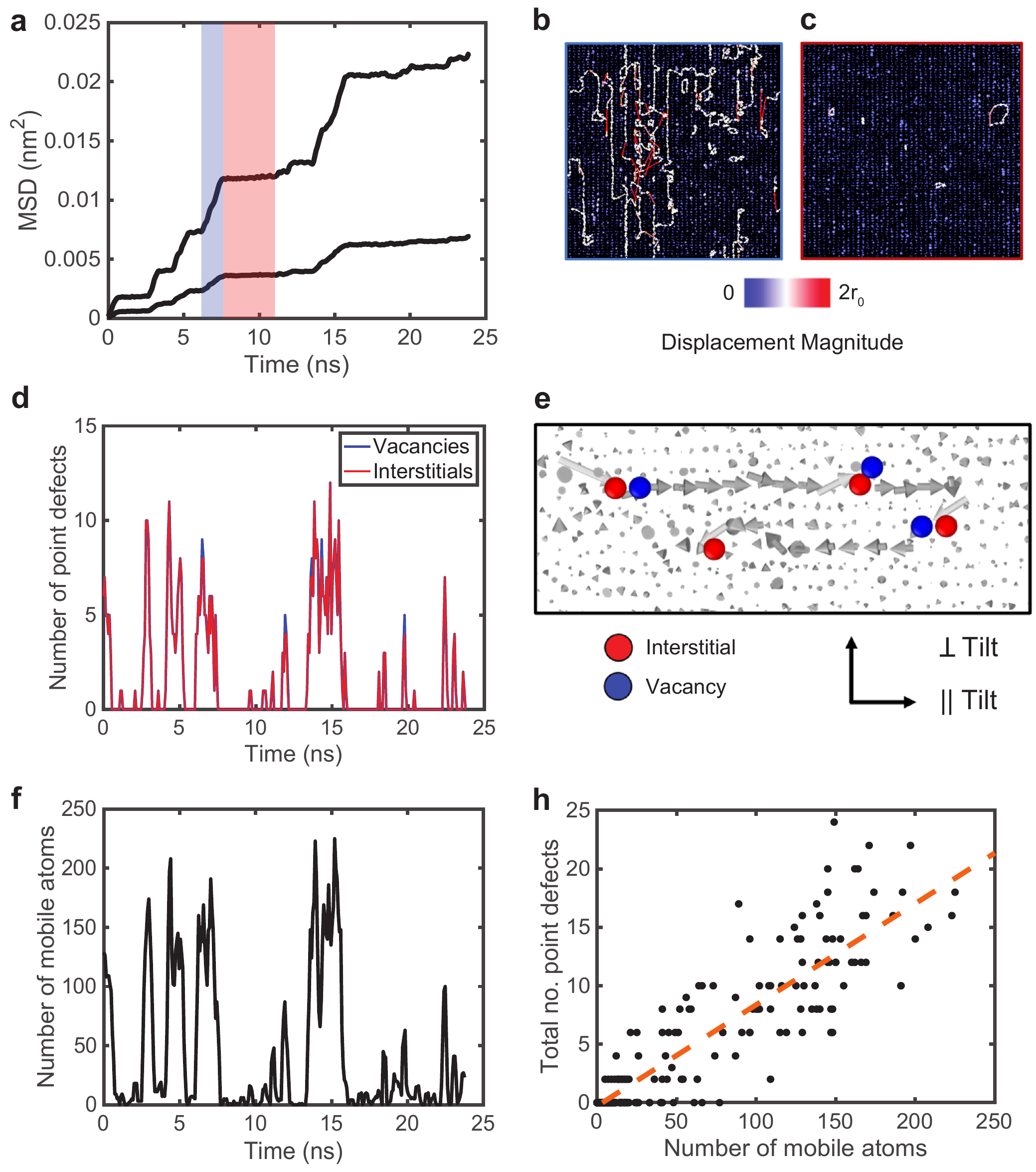}
\caption{Intermittent diffusion in the $\Sigma3$ GB for a typical GB cross-section
(10 nm $\times$ 10 nm). (a) Example of atomic MSD as a function of
time at 710 K (0.77 $T_{m}$). (b,c) Typical atomic displacement diagrams
during the (b) avalanche and (c) locked time periods corresponding
to the blue and red regions in (a). The arrows represent net displacements
during time intervals of 1.2 ns and 3 ns and are colored by the displacement
magnitude. Note the enhanced atomic mobility in the avalanche regime
and only a few displaced atoms in the locked regime. (d) Typical collective
diffusion event involving Frenkel pairs. The arrows represent string-like
displacements during the time $\Delta t=80$ ps colored in dark grey
and light grey for displacements smaller and larger than the first-neighbor
distance $r_{0}$. (e) Peaks in the numbers of vacancies and interstitials
correlate with the locked and avalanche time periods seen in (a).
(f) Peaks in the number of mobile atoms also correlate with the locked
and avalanche time periods. (h) Correlation between the point-defect
population and the number of mobile atoms in the GB core with the
correlation factor of $R^{2}=0.83$.}
\label{fig:fig3}
\end{figure}

Our first key observation is the significant intermittency of GB diffusion
at the intermediate temperatures. The effect is illustrated in Fig.~\ref{fig:fig3}a,
where we plot the mean-square displacement (MSD) of atoms in the $\Sigma3$
GB core as a function of time. The remarkable feature of the plot
is the existence of the slope fluctuations with stochastic switches
between nearly horizontal and large-slope portions. The zig-zag shape
of this plot points to the existence of two dynamic regimes with a
very small and a relatively large GB diffusivity. We refer to the
plateaus of the diffusivity as the \emph{locked} states and to the
surges of diffusion as \textit{avalanches}. The times at which the
avalanches nucleate correspond to abrupt increases in the number of
mobile atoms in the GB core, as illustrated in Fig.~\ref{fig:fig3}b.
In contrast, during the locked periods, most GB atoms remain immobile;
only a few atoms execute jumps, often in the form of small rings of
highly correlated displacements (Fig.~\ref{fig:fig3}c).

The diffusivity bursts are dynamic signatures of the avalanches, but
they also have a structural signature. Each avalanche period is accompanied
by a drastically increased point defect activity in the GB core and
thus a temporary decrease in the structural order. An avalanche triggers
a massive formation of vacancies and interstitials, which occur in
nearly equal amounts. Comparison of Figs.~\ref{fig:fig3}a and \ref{fig:fig3}e
shows that the numbers of vacancies and interstitials in the GB core
strongly correlate not only with each other but also with the occurrence
of locked and avalanche time periods. The fluctuating point defect
populations also correlate with the total number of mobile atoms (Fig.~\ref{fig:fig3}f,g),
confirming that GB diffusion occurs by a defect-mediated mechanism.
At 700 K, a typical avalanche creates hundreds of point defects and
activates thousands of mobile atoms.

The questions then arise: how does an avalanche begin, and how does
it end? Careful structural analysis (see Methods) reveals that a typical
avalanche starts with the formation of an interstitial atom in the
GB core that hops multiple times (on the ps timescale) before pushing
a regular atom out of its preferred site (Fig.~\ref{fig:fig3}h).
This event triggers a cascade of collective structural excitations
with a string-like character in which multiple atoms move simultaneously
in the same direction. A string-like excitation can be viewed as an
extended Frenkel pair in which the leading atom has an interstitial
character (associated with local compression) and the trailing atom
has a vacancy character (local tension). This interpretation is confirmed
by analyzing the free volume distribution in the population of mobile
atoms (see Supplementary Fig.~1). Importantly, the avalanches involve
the \emph{facilitated dynamics} \citep{jung2005dynamical,chandler2010dynamics}
of the multiple Frenkel pairs and neighboring string-like excitations.
Namely, a collective atomic displacement at one location triggers
nearby diffusive events by creating a positive feedback for further
propagation of the atomic displacement field that rapidly grows in
size into a full-fledged avalanche. As the point-defect population
grows, so does the probability of their mutual recombination. Eventually,
a bifurcation point is reached at which the dynamics switch to a recombination
regime. We hypothesize that the locked states arise from such spontaneous
switches of the diffusion process from the multiplication to the recombination
regime. Self-annihilation of point defects across the periodic boundary
conditions can also play a role.

\begin{figure}[H]
\centering\leavevmode \includegraphics[width=0.43\textwidth]{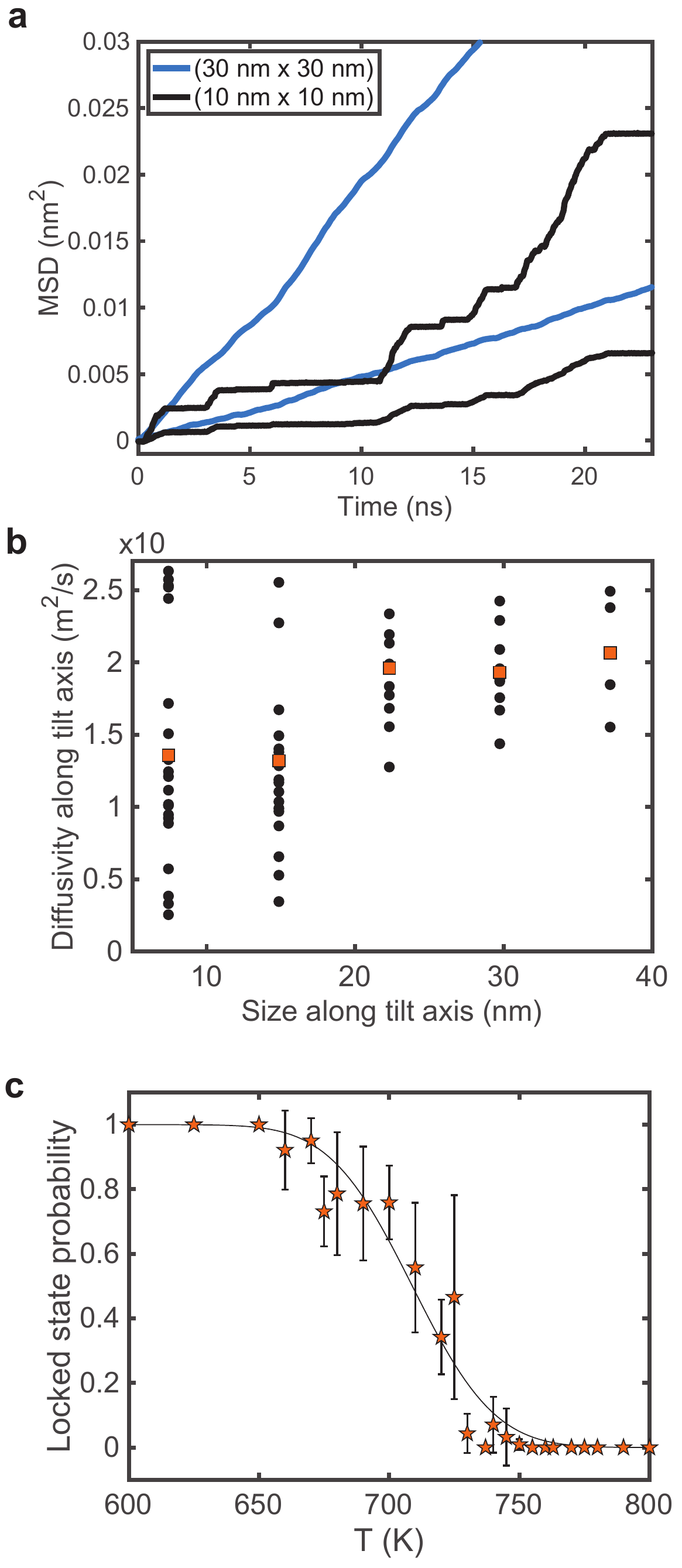}
\caption{Size dependence of diffusion in the Al $\Sigma3$ GB. (a) Atomic MSD
versus time at the temperature of 700 K for the (10 nm $\times$ 10
nm) and (30 nm $\times$ 30 nm) GB cross-sections. Note that the diffusion
intermittency disappears with increasing cross-section. (b) Size dependence
of GB diffusivity for several different cross-sections with a fixed
13 nm size perpendicular to the tilt axis and a varied size from 6
nm to 38 nm along the tilt axis. The points represent independent
MD runs with different velocity seeds. The red points represent the
average values. (c) Locked-state probability (fraction of plateau
portions in (a)) as a measure of intermittency plotted as a function
of temperature. The error bars are obtained by using up to ten velocity
seeds at each temperature. The sigmoidal fit is only meant to guide
the eye. Note the suppression of intermittency with increasing temperature.}
\label{fig:fig6}
\end{figure}

Our second key finding is that the intermittent behavior of GB diffusion
is system-size dependent on the $\sim10$ nm length scale. This was
demonstrated by increasing the GB cross-section and remeasuring the
diffusion coefficients and the MSD curves. In the example shown in
Fig.~\ref{fig:fig6}a, the initial zig-zag shape of the MSD curves
is replaced by a linear function when the cross-section is increased
from (10 nm $\times$ 10 nm) to (30 nm $\times$ 30 nm). In a more
systematic test, the GB cross-section in the direction parallel to
the tilt axis was increased incrementally while keeping the size normal
to the tilt axis fixed. For each cross-section, the diffusivity was
measured several times with different velocity seeds as shown in Fig.~\ref{fig:fig6}b.
The scatter of the points at each cross-section is a measure of the
intermittent diffusion behavior. The plot clearly shows that the scatter
decreases with increasing GB size. (The average diffusivity values
seem to increase with the size, but this trend does not pass statistical
significance tests.) The probability of locked states can be used
as another measure of intermittency. The plot Fig.~\ref{fig:fig6}c
shows that, for a fixed (10 nm $\times$ 10 nm) cross-section, the
intermittency decreases with increasing temperature and vanishes at
about 750 K. These tests demonstrate that the intermittent behavior
of GB diffusion exists in a cartain temperature-size domain and disappears
at high temperatures and/or large GB sizes.

We posit that both phenomena, the GB diffusion intermittency and its
size dependence, have the same physical origin: the finite size of
the mobile atom clusters generated by the point-defect avalanches.
In a large GB area, multiple avalanches can form at different moments
of time at random locations. The measured diffusion coefficient then
represents the atomic mobility averaged over multiple avalanches and
exhibits a smooth and size-independent behavior. If the GB size is
less than or comparable to the avalanche size, then the boundary conditions
truncate the mobile cluster size to the GB size. Each mobility surge
then quickly spreads over the entire GB area and later stops across
the entire GB area, causing the on-and-off behavior seen in Figs.~\ref{fig:fig3}a
and \ref{fig:fig6}a. Due to the anisotropy of GB structure, the intermittency
condition can be met when the mobile cluster size becomes comparable
to the GB dimension in at least one direction. The disappearance of
the diffusion intermittency at high temperatures is explained by a
decrease in the mobile cluster size with temperature.

To provide mechanistic support to this explanation, we have analyzed
the dynamic length scales of the GB diffusion in more detail. We find
that the measures of dynamic length scales used in previous work \citep{gbmdhpnas,AlSm1}
predict lengths that are too small in magnitude to explain the observed
diffusion intermittency. For example, the average (1.4 nm) and maximum
(7 nm) string lengths computed for the $\Sigma3$ GB at 700 K by the
standard string-like segmentation method (see Methods) are smaller
than the observed $\sim10$ nm system-size effect. A method capturing
the length scale of the entire avalanche region was required.

\begin{figure}[H]
\noindent \begin{centering}
\includegraphics[width=0.97\textwidth]{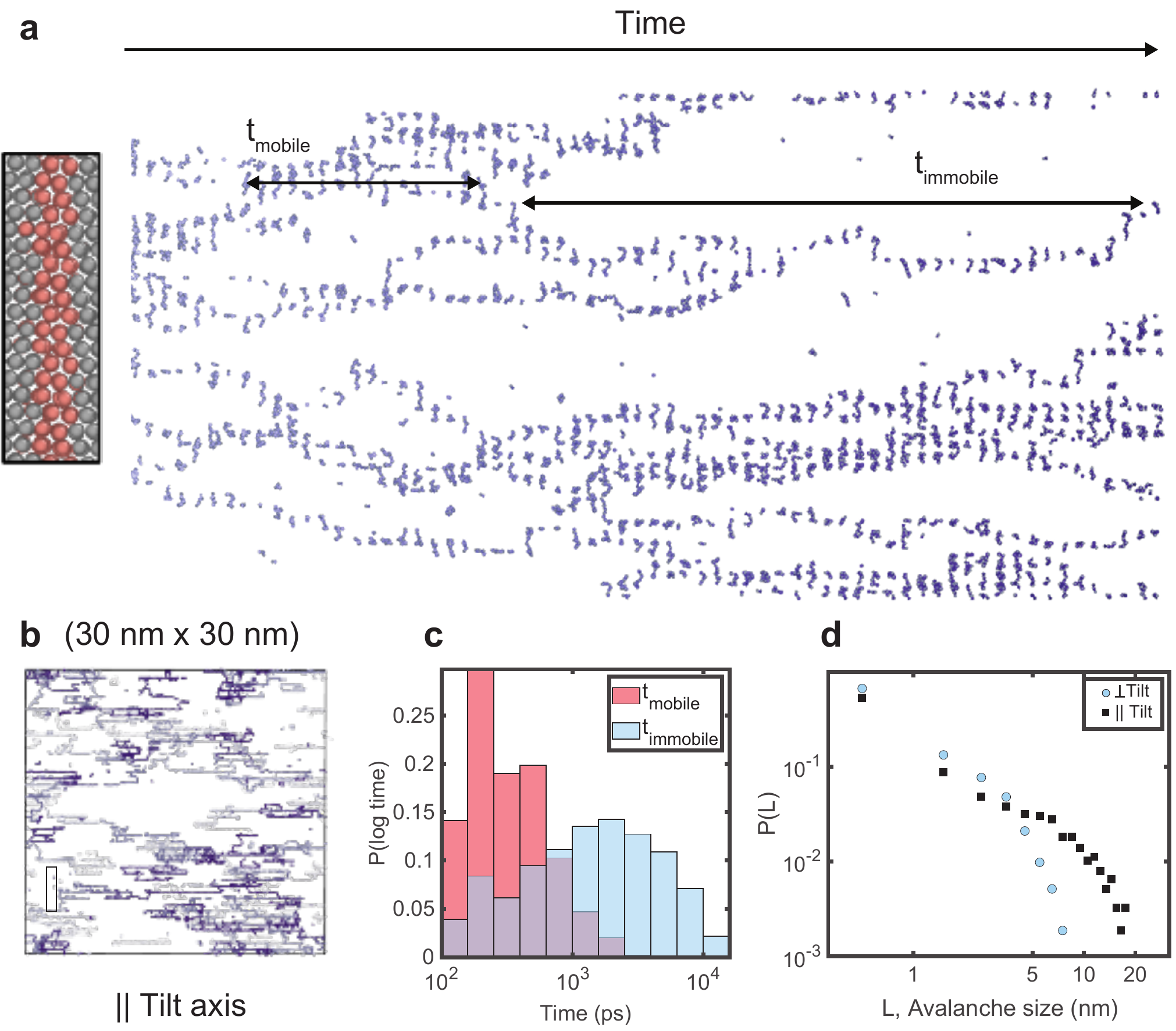}
\par\end{centering}
\caption{Space-time diagrams showing dynamic facilitation of GB diffusion.
(a) Space-time atomic mobility diagram for the $\Sigma3$ GB at 700
K computed with the time interval of $\Delta t=40$ ps and viewed
perpendicular to the tilt axis. The GB structure is shown on the left
with light grey FCC atoms and dark grey non-FCC atoms. The immobile
clusters are evident as voids (``bubbles'') \citep{chandler2010dynamics}
between the mobile atom trajectories. The arrows show examples of
mobile and immobile cluster dimensions in time. (b) Cross section
of the same diagram with all time data projected onto the GB plane.
The rectangular box indicates the 6 nm scale of the GB structure shown
in (a) relative to the GB area. (c) Probability distribution of lifetimes
of mobile and immobile clusters. Note that immobile clusters persist
on a longer time scale than mobile clusters. (d) Avalanche size distribution
at 720 K as measured by the maximum avalanche size perpendicular and
parallel to the tilt axis. The avalanches are elongated along the
tilt axis and exhibit a cutoff behavior with large avalanches becoming
increasingly unlikely with increasing size. Note that the cutoff size
along the tilt axis is larger than 10 nm. \label{fig:(a)-facilitation}}
\end{figure}

As stated above, the avalanches exist due to the dynamic facilitation
of collective atomic displacements, in which a local excitation spreads
in a self-catalytic manner by increasing the probability of atomic
displacements in neighboring regions. Facilitated dynamics can be
conveniently visualized using space-time diagrams \citep{jung2005dynamical}.
To construct such a diagram, atomic displacement fields corresponding
to a fixed time interval are stacked together along the time axis
and viewed in a space-time cross-section (Fig.~\ref{fig:(a)-facilitation}a,b).
The diagram reveals the formation and disappearance of mobile and
immobile clusters in space and time. The GB structural units blink
between on and off states corresponding to the diffusive and quiescent
time periods. The branching and coalescence of diffusion trajectories
on the diagram are signatures of facilitated dynamics. We have developed
a space-time clustering algorithm (see Methods) that identifies the
mobile and immobile clusters and measures their dimensions in space
and time. The time dimensions of these clusters are associated with
the lifetimes of the avalanches and locked states. As demonstrated
in Fig.~\ref{fig:(a)-facilitation}c, we find that immobile clusters
have a longer average lifetime than mobile clusters. This time scale
separation is a known hallmark of facilitated dynamics \citep{jung2005dynamical,chandler2010dynamics}.
The spatial dimensions of the mobile and immobile clusters are measures
of avalanche size. Fig.~\ref{fig:(a)-facilitation}d shows the avalanche
size probability distribution computed for the $\Sigma3$ GB at 700
K. The plot reveals that the avalanches are elongated parallel to
the tilt axis (which is consistent with the diffusion anisotropy)
and have a significant probability of being 5 to 15 nm in size. Over
15 percent of the mobile atoms participate in avalanches larger than
10 nm along the tilt axis. The largest avalanches contain a thousand
or more atoms and have linear dimensions up to 20 nm along the tilt
axis. These dimensions are well-consistent with the length scale of
the system-size effect of GB diffusion (cf.~Figs.~\ref{fig:fig3}a
and \ref{fig:fig6}a). That the avalanche size decreases with temperature
was confirmed by the analysis shown in Supplementary Fig.~2, in which
the average avalanche size was estimated by the virtual system-size
analysis explained in Methods.

All simulations discussed so far utilized the Al-M interatomic potential
\citep{Almendelev}. To demonstrate the generality of our results,
selected simulations were repeated with two additional potentials:
the embedded-atom potential from Ref.~\citep{Al99} (Al-99) and the
recently developed physically-informed neural network (PINN) potential
\citep{AlPINN2} referred to here as Al-PINN. The latter is computationally
much slower than the Al-M and Al-99 potentials but provides the most
accurate description of Al properties. An Arrhenius diagram summarizing
the diffusion coefficients in the $\Sigma3$ and $\Sigma17$ GBs computed
with these potentials is shown in Supplementary Fig.~3a. All three
potentials display similar diffusion behaviors, with a comparable
degree of diffusion anisotropy and similar temperatures of premelting.
More importantly, the existence of the collective diffusion mechanisms
in the $\Sigma3$ and $\Sigma17$ GBs is reproduced with all three
potentials. In particular, the Al-PINN simulations predict string-like
displacement geometries visually indistinguishable from those obtained
with the Al-M potential. Furthermore, simulations with the Al-99 potential
revealed the same type of diffusion intermittency as observed with
the Al-M potential (see example in Supplementary Fig.~3b). The intermittent
diffusion behavior and its disappearance with increasing system size
were also confirmed for the (100) twist and HAIC boundaries, as illustrated
in Fig.~\ref{fig:fig6}. Although these boundaries are crystallographically
and structurally different from the $\Sigma3$ and $\Sigma17$ GBs,
they exhibit the same correlation between the avalanche time periods
on the MSD plots and the bursts of point-defect activity as in the
$\Sigma3$ and $\Sigma17$ GBs. 

To further demonstrate the generic character of our main results and
conclusions, we performed additional diffusion simulations for the
$\Sigma3$ GB in FCC Ni and $\Sigma17$ GB in BCC Fe, see Supplementary
Table 1. We again observed the diffusion intermittency on the 10 nm
scale at intermediate temperatures, with the avalanche periods associated
with bursts of point defect generation (see Supplementary Figure 4).
We note that collective diffusion mechanisms were previously studied
in BCC Fe GBs in \citep{starikov2020study}, although diffusion intermittency
was not documented. The observation of diffusion avalanches across
multiple GB structures and chemistries provides strong evidence that
the avalanche-mediated diffusion behavior reported here is generic
to metallic GBs.

\section*{\textcolor{black}{Discussion}}

The previous paradigm of GB diffusion has been that this process is
continuous in time and uniform across the GB area. Contrary to this
idealistic picture, we have shown that there is a temperature-size
domain in which GB diffusion is discontinuous in both time and space.
For the metallic GBs studied here, the time and space scales of the
diffusion intermittency are several nanoseconds and several nanometers,
respectively, although the exact numbers are system-dependent. The
physical origin of the GB diffusion intermittency is the dynamic heterogeneity
of the diffusion process. The atomic displacements are not distributed
uniformly along the boundary but cluster together into large groups
of mobile atoms undergoing a collective rearrangement. Spontaneous
formation of a mobile cluster is triggered by a local atomic rearrangement
that spreads like an avalanche by the dynamic facilitation mechanism.
The avalanche region is characterized by a drastically increased defect
population and enhanced atomic mobility. If the avalanche size is
comparable to the GB size in at least one direction, a spike in the
GB diffusivity arises. This is the cause of the system-size dependence
of GB diffusion and the reason why the size effect vanishes when the
GB area becomes large.

To put our findings in perspective with the literature, dynamic heterogeneity
has been studied for years, but mostly in disordered bulk systems
such as supercooled liquids and glasses \citep{donati1998stringlike,derlet2021viscosity,AlSm1,annamareddy_superionic}.
The concept of dynamic facilitation was introduced more recently \citep{jung2005dynamical,pan2005heterogeneity,chandler2010dynamics}
but again in the context of bulk disordered systems. Zhang et al.~\citep{gbmdhpnas}
have shown that GB migration and diffusion processes exhibit dynamic
heterogeneity similar to that in supercooled liquids. They also observed
string-like atomic displacements on time and length scales similar
to those in supercooled liquids. However, their work was focused primarily
on GB migration rather than diffusion. Dynamic heterogeneity and collective
diffusion mechanisms were also investigated in highly disordered,
premelted copper GBs \citep{mishin2015atomistic}. Intermittent diffusion
behavior was seen in MD simulations of low-angle twist GBs, but no
explanation was attempted \citep{schonfelder2005comparative}. None
of the prior work \citep{gbmdhpnas,mishin2015atomistic,schonfelder2005comparative}
investigated the physical origin of the GB diffusion intermittency
or its size effect. On the other hand, avalanche-mediated dynamics
were investigated by MD simulations of several processes, such as
crystallization of glasses (see, e.g, \citep{Eduardo:2014wl} and
references therein), but not for GB or even bulk diffusion. It should
also be noted that the avalanches during glass crystallization occur
in a metastable state as the system gradually transitions to a more
stable state. By contrast, the avalanches discussed in this work occur
in a fully equilibrium system and are caused by thermal fluctuations
away from equilibrium. Thus, although some of the individual components
of this work pre-existed, they were expanded and integrated into a
coherent mechanistic explanation of the previously unknown effects
reported in this article.

One lesson from this work for future atomistic simulations is that
the system size effect is important. If the GB size is comparable
to or smaller than the avalanche size and the simulation time is too
short, the MSD-vs-time curve can represent either an avalanche or
a locked state. In either case, the GB diffusivity extracted from
the Einstein relation will be far from the real value. Multiple simulations
with different system sizes and diffusion times are required to ensure
averaging over multiple avalanche-locked cycles to obtain the converged
GB diffusivity. In experiments, a nano-scale GB can exhibit intermittent
mass transport, creating an additional source of noise in electronic
devices. The situation can, for example, be relevant to mass diffusion
and/or electromigration in conductor lines in integrated circuits.
Assuming a nanosecond time scale of the avalanches, the respective
fluctuation frequency could be in the GHz range.

In addition to the practical aspects mentioned above, this work raises
many fundamental questions. In contrast to the homogeneously disordered
matter such as 2D or 3D supercooled liquids and glasses, GBs are partially
ordered quasi-2D systems with a tunable structure and structural anisotropy.
They can support a broad spectrum of local structural excitations
(defects) with diverse sizes, energetics, and geometries. The effects
of this partial order and the structural richness on the dynamic heterogeneity
require further investigations. The exact mechanism of the dynamic
facilitation in GBs remains unknown but could be different from that
in homogeneously disordered systems. For example, we found a much
larger avalanche length scale (up to $\sim70r_{0}$) compared to that
in supercooled liquids and glasses (below $\sim10r_{0}$) \citep{royall2015role}
($r_{0}$ being the first nearest neighbor distance). Furthermore,
this work only studied GBs in single-component systems. Adding segregating
solutes to the GB core will add a new level of complexity but will
also provide a means to control the GB dynamics. While the segregation
effect on GB diffusion has been studied \citep{PhysRevMaterials.4.073403},
the underlying microscopic mechanisms remain unknown.

Finally, even though most GB diffusion measurements are performed
on time and space scales far exceeding the avalanche scales, one should
explore alternative methods capable of extracting at least some mechanistic
information. For example, isotope effect measurements provided evidence
for collective diffusion mechanisms in metallic glasses \citep{faupel2003diffusion}.
Although such measurements do not give detailed information about
specific mechanisms and are limited to elements with available isotopes,
their application to GB diffusion could still be attempted to confirm
the existence of collective atomic rearrangements predicted by the
simulations. Dielectric spectroscopy methods have also been used in
specific glasses and supercooled liquids to resolve the characteristic
frequencies of slow and fast processes during glassy dynamics \citep{lunkenheimer2002dielectric,jug2021structural,biroli2022statistical}.
They, too, could potentially be used to gain at least some insights
into the GB diffusion dynamics on the atomic level.

\section*{\textcolor{black}{Methods}}

\noindent The Large-scale Atomic/Molecular Massively Parallel Simulator
(LAMMPS) \citep{LAMMPS} was used to conduct molecular statics and
MD simulations. The software package OVITO \citep{OVITO} was used
for visualization and analysis of the collective diffusion mechanisms.\textcolor{black}{{}
The interatomic potentials were downloaded from the LAMMPS potential
library. The GitHub codes from~\citep{PINN-Github} were used for
LAMMPS simulations with the Al-PINN potentia}l.

Unrelaxed Al GB structures were created by joining two grains in a
simulation box with periodic boundary conditions in the GB plane ($x$-$y$
coordinate plane) and free-surface boundary conditions normal to the
GB ($z$-direction). The GB cross-section was chosen with dimensions
at least 10 nm by 10 nm in the $x$ and $y$ directions, and each
grain had a thickness of at least 6 nm. To study finite-size effects
on diffusion, several GB cross-sections were tested up to 32 nm by
32 nm. For the commensurate GBs (Table \ref{tab:1}), the lateral
bicrystal dimensions were chosen with integer repeats of the coincident
site lattice (CSL) unit cell. For the incommensurate GB, the misfit
strain was minimized within the maximum interface size (30 nm by 30
nm) and applied to the upper grain only. The minimization procedure
was repeated at different temperatures for lattice constants expanded
by the precomputed bulk thermal expansion coefficients. The maximum
absolute value of the misfit strain used in this work was 1.001.

The initial GB structures were optimized by a standard grid search
that seeks to find a deep local energy minimum \citep{olmsted2009survey,tschopp2015symmetric}.
Several hundred initial unrelaxed GB structures were prepared with
different rigid shifts applied to the upper grain relative to the
lower. Translations were chosen within a unit cell of the reciprocal
lattice of the CSL (known as the displacement shift complete lattice)
and included overlaps of the two grains. For each initial structure,
pairs of closely spaced atoms were identified and one atom was deleted
at random if the overlap radii were in the range 0.7-0.95$r_{0}$,
where $r_{0}=0.28$ nm is the first nearest neighbor distance in FCC
Al. Conjugate gradient minimization was performed at 0 K to relax
the atoms in the GB core of each candidate structure. The lowest energy
structure was used as input for the subsequent diffusion simulations.
This optimization procedure recovered the well-known structures of
the $\Sigma3$ and $\Sigma17$ GBs, but it did not sample different
GB densities as in more sophisticated optimization routines \citep{frolov2013structural,hickman2017extra,yang2020grain}.

Wigner-Seitz analysis as implemented in OVITO was used to analyze
the point defect content of GBs. This method identifies point defects
in a finite-temperature GB structure relative to the Voronoi tessellation
of a defect-free reference structure. As the reference state, we used
the 0 K GB structure homogeneously expanded by the thermal expansion
coefficient at the chosen temperature.

GB diffusion was studied by canonical (NVT) and microcanonical (NVE)
MD simulations. A multi-step equilibration procedure was performed
before production runs. First, the bicrystal was homogeneously expanded
by the precomputed bulk thermal expansion strain at the chosen temperature.
Next, a 1 nm thick rigid slab at the top of the upper grain was allowed
to float in $z$-direction during a short (0.1 ns) NVT anneal. Combined
with the pre-expansion, this step reduced the pressure in the grains
to near zero. Then, the rigid slabs at the top and bottom of the bicrystal
were fixed and an NVT anneal was performed for 1 ns. The fixed boundary
conditions suppressed spontaneous GB sliding events allowing us to
clearly identify the diffusion mechanisms. After the equilibration,
production NVT runs were performed for up to 40 ns and several hundred
dump files were output for the subsequent calculation of the diffusion
coefficients. NVE simulations were then performed starting from the
end of each NVT run to analyze the diffusion mechanisms. The NVE ensemble
was chosen to eliminate any possible effect of the thermostat on spontaneous
diffusive events. For selected calculations, including diffusion measurements
at low temperatures, additional statistics were obtained by performing
multiple runs with different velocity seeds.

The GB self-diffusion coefficients were measured by tracking atoms
within a 2 nm thick layer centered at the GB. The current GB position
was determined as the average $z$-coordinate of non-FCC atoms identified
by the polyhedral template matching modifier in OVITO \citep{larsen2016robust}.
The diffusion coefficients within the layer were calculated from the
Einstein relations $D_{x}=\langle x^{2}\rangle/2t$ and $D_{y}=\langle y^{2}\rangle/2t$,
where $\langle x^{2}\rangle$ and $\langle y^{2}\rangle$ are the
MSDs of atoms in the respective directions parallel to the GB plane
and $t$ is the simulation time. Bootstrap resampling was employed
as in \citep{race2015quantifying} to compute error estimates associated
with the diffusion coefficient. The hyper-parameters chosen for this
method include a smoothing window of 5 ps, a block length of 20 ps,
and the number of resampled trajectories equal to 100. Note that using
a fixed-width layer can underestimate the diffusion coefficient because
of the possible inclusion of some of the immobile atoms from the perfect
lattice regions next to the GB core. This underestimation was corrected
by rescaling the diffusion coefficients by the average inverse fraction
of the mobile atoms in the layer at long times. The convergence of
diffusion calculations in time was verified by checking that the slope
of the log(MSD)-log($t$) plot was equal to one. All diffusion coefficients
reported here satisfied this condition.

At high temperatures (typically, $T>0.85T_{m}$) and depending on
the GB, random interface displacements were observed in addition to
diffusion. When the average interface position changed by more than
0.5 nm, the diffusion calculation was reset from the new GB position.

To compute the liquid diffusion coefficients, an initial 32,000-atom
liquid structure was created in a periodic box by heating a perfect
Al crystal in the isothermal-isobaric (NPT) ensemble to 1500 K and
holding it for 200 ps. Next, a set of liquid samples was generated
in the temperature range from 1050 to 1450 K by quenching the high-temperature
structure at a rate of 50 K/100 ps and holding it at the set temperature
for 200 ps. A further stepwise quenching was performed from 1050 K
to 600 K at a rate of 25 K/ns with a hold time of 2 ns every 50 K.
These times were sufficient for liquid equilibration at the chosen
temperatures as evidenced by converged total potential energy and
linear MSD vs time behavior. Crystallization was observed at and below
650 K, and these temperatures were not used for the diffusion calculations.
Production NVE anneals were performed for all liquid structures for
0.1 ns. The diffusivity was computed from the 3D Einstein relation
$D=\langle r^{2}\rangle/6t$ and averaged over 20 independent runs
with different velocity seeds.

The following algorithm was used to reveal string-like displacements
in GB diffusion. First, mobile atoms were identified within the GB
core such that the net displacement of an atom $\Delta r$ during
a preset time interval $\Delta t$ lay within the range $0.4r_{0}<\Delta r<1.2r_{0}$.
Here, the upper bound was chosen to eliminate atoms that had undergone
multiple hops, and the lower bound was chosen to eliminate immobile
atoms. Next, mobile atomic pairs $(i,j)$ were found that remained
nearest neighbors at the times $t=0$ and $t=\Delta t$ and satisfied
the criterion $\text{min}(|r_{i}(t)-r_{j}(0)|,|r_{j}(t)-r_{i}(0)|)<0.43r_{0}$.
This criterion captures atomic pairs with string-like motion in which
one atom jumps into the previous position of the other. The algorithm
has three hyper-parameters: the lower and upper bounds for the displacement
and the substitution distance (the numbers $0.4$, $1.2$ and $0.43$
above). These parameters were chosen based on prior studies of diffusion
in glass-forming supercooled liquids \citep{gbmdhpnas,AlSm1} and
were fixed across all GBs studied in this work.

All mobile pairs containing common atoms were then joined into larger
string-like clusters, and each cluster was assigned a unique ID. Cluster
statistics were then computed over all clusters, including the number
of atoms and the gyration tensor. The string identification algorithm
was repeated for a range of time intervals $\Delta t$ with multiple
start times. The average string length, measured by the average number
of atoms in the string-like cluster, typically exhibits a single maximum
as a function of $\Delta t$ at a characteristic time $\Delta t=t_{s}^{*}$.

The analysis of diffusion avalanches required a more flexible definition
of mobile clusters. The input to this method is the space-time trajectory
consisting of atomic displacement fields computed at a fixed time
interval ($\Delta t$) and stacked in time. Here, $\Delta t$ is a
hyper-parameter taken to be approximately $0.02t_{s}^{*}$. First,
the string clustering step was performed in space-time with no upper
bound on the displacement length and all other hyper-parameters as
above. The absence of an upper bound allows the strings to incorporate
atoms executing multiple jumps. Next, the string connectivity criterion
was relaxed: neighboring strings were combined in space if at least
two atoms in adjacent strings (either in time or space) were no more
than $1.2r_{0}$ apart. This hyper-parameter can be interpreted as
a cutoff distance for the interaction range in dynamic facilitation.
It also provides a good visual segmentation of the data. If the cutoff
distance is too large, or the trajectories in space-time are too dense,
then all atoms are part of the same mobile cluster and the algorithm
loses its meaning. In the space-time trajectories, the same atom can
be included multiple times in the same mobile cluster if it executes
multiple jumps. The mobile clusters often form elongated, anisotropic
shapes in space-time. Their size was quantified as the number of unique
atoms in the cluster. The dynamic length scale of a given mobile cluster
was quantified as its maximum width along a given direction in the
GB plane, and its timescale as the maximum width along the time axis.

Immobile clusters are apparent in space-time diagrams as empty regions
(``bubbles'' \citep{chandler2010dynamics}) separating the mobile
clusters. In principle, their length and time scales can be computed
in a similar manner as for the mobile clusters. However, in practice
a single percolating immobile cluster in space-time is often found
with this method. An alternative method used here defines and computes
the immobility time by considering individual square ($l\times l$)
GB patches in space. The immobility time in each patch is then computed
as the time between mobility events. The immobility time scale is
obtained in the limit of $l$ approaching zero, while the immobility
time distribution can be computed at $l=r_{0}$. The ratio of the
immobility and mobility times can also serve as a measure of diffusion
intermittency. Considered as a function of the patch size $l$, this
ratio is used in the virtual size-effect tests as shown in Supplementary
Fig.~2.

\noindent \textcolor{black}{\bigskip{}
}

\noindent \textbf{\textcolor{black}{Data availability.}}\textcolor{black}{{}
The data supporting the findings of this study are available in the
Supplementary Information file or from the corresponding authors upon
reasonable request. The computer simulation part of this research
used the publicly available codes LAMMPS and OVITO. The routine computer
scripts controlling the execution of the calculations are not central
to this work but are available from the corresponding authors upon
reasonable request.}

%\textcolor{black}{\bibliographystyle{ActaMatnew}
%\bibliography{citations}
%}

\noindent \textcolor{black}{\bigskip{}
}

\noindent \textbf{\textcolor{black}{Acknowledgements}}\textcolor{black}{{}
}\\
\textcolor{black}{We would like to thank Raj Koju, Jack Douglas and
Peter Derlet for helpful discussions. This research was supported
by the National Science Foundation, Division of Materials Research,
under Award no.~2103431.}

\noindent \textcolor{black}{\bigskip{}
}

\noindent \textbf{\textcolor{black}{Author contributions}}\textcolor{black}{{}
}\\
\textcolor{black}{Y.M.~acquired funding for this research. I.C. wrote
all computer codes/scripts required for this work and conducted all
simulations reported here under Y.M.\textquoteright s guidance and
supervision. Both authors participated equally in interpreting the
results and writing the paper, and approved its final version.}

\noindent \textcolor{black}{\bigskip{}
}

\noindent \textbf{\textcolor{black}{Additional information}}\textcolor{black}{}\\
\textcolor{black}{Supplementary Information accompanies this paper
at }\texttt{\textcolor{black}{https://doi.org/...}}

\noindent \textcolor{black}{\bigskip{}
}

\noindent \textbf{\textcolor{black}{Competing interests:}}\textcolor{black}{{}
The authors declare no competing interests.}

\textcolor{black}{\newpage\clearpage}

\begin{table}[H]
\centering %
\begin{tabular}{lcccc}
\toprule 
GB description & $x$ & $y$ & $z$ & $\gamma$ (mJ/m$^{2}$)\tabularnewline
\midrule 
\multirow{2}{*}{$\Sigma3$ $\left\langle 110\right\rangle $ tilt 70.5$^{\circ}$} & $[-1,1,1]$ & $[0,1,-1]$ & $[2,1,1]$ & \multirow{2}{*}{418}\tabularnewline
 & $[1,-1,-1]$ & $[0,-1,1]$ & $[2,1,1]$ & \tabularnewline
 &  &  &  & \tabularnewline
\multirow{2}{*}{$\Sigma17$ $\left\langle 100\right\rangle $ tilt 61.9$^{\circ}$} & $[5,-3,0]$ & $[0,0,-1]$ & $[3,5,0]$ & \multirow{2}{*}{488}\tabularnewline
 & $[-5,3,0]$ & $[0,0,1]$ & $[3,5,0]$ & \tabularnewline
 &  &  &  & \tabularnewline
\multirow{2}{*}{$\Sigma85$ $\{100\}$ twist 8.8$^{\circ}$} & $[-1,13,0]$ & $[13,1,0]$ & $[0,0,1]$ & \multirow{2}{*}{306}\tabularnewline
 & $[1,13,0]$ & $[-13,1,0]$ & $[0,0,1]$ & \tabularnewline
 &  &  &  & \tabularnewline
\multirow{2}{*}{HAAI \{100\}$||$\{111\}} & $[1,1,0]$ & $[1,-1,0]$ & $[0,0,1]$ & \multirow{2}{*}{306}\tabularnewline
 & $[1,1,-2]$ & $[1,-1,0]$ & $[1,1,1]$ & \tabularnewline
\bottomrule
\end{tabular}\caption{Crystallography and 0 K energies ($\gamma$) of GBs studied in this
work. The Cartesian $x$, $y$ and $z$ axes are aligned with edges
of the rectangular simulation box. The GB plane is normal to the $z$-direction.
For each GB, the two lines indicate crystallographic directions parallel
to the Cartesian axes in the upper and lower grains. The GB energy
was computed with the interatomic potential from Ref.~\citep{Almendelev}.\label{tab:1}}
\end{table}

\begin{table}[H]
\centering %
\begin{tabular}{lccc}
\toprule 
GB description & Direction & $E$ (eV) & $D_{0}$ (m$^{2}$/s)\tabularnewline
\midrule 
$\Sigma3$ $\left\langle 110\right\rangle $ tilt 70.5$^{\circ}$ & $\parallel$ tilt axis & 1.31 $\pm$ 0.03 & (3.89 $\pm$ 7.8)$\times10^{-3}$\tabularnewline
 & $\perp$ tilt axis & 1.45 $\pm$ 0.03 & (1.17 $\pm$ 2.24)$\times10^{-2}$\tabularnewline
$\Sigma17$ $\left\langle 100\right\rangle $ tilt 61.9$^{\circ}$ & $\parallel$ tilt axis & 0.72 $\pm$ 0.01 & (5.53 $\pm$ 1.59)$\times10^{-6}$\tabularnewline
 & $\perp$ tilt axis & 0.81 $\pm$ 0.01 & (2.09 $\pm$ 0.74)$\times10^{-5}$\tabularnewline
Supercooled liquid &  & 0.227 $\pm$ 0.001 & (9.69 $\pm$ 0.24)$\times10^{-8}$\tabularnewline
\midrule 
FCC crystal: &  &  & \tabularnewline
Experiment \citep{Zhong:2020we} &  & 1.32 & 1.79 $\times10^{-5}$\tabularnewline
DFT calculations \citep{Hooshmand:2020tu} &  & 1.25$-$1.35 & 5.42 $\times10^{-6}$$-$2.42 $\times10^{-5}$\tabularnewline
\bottomrule
\end{tabular}\caption{Activation energies ($E$) and prefactors ($D_{0}$) calculated in
this work. The reference data for diffusion in FCC Al is taken from
experiments \citep{Zhong:2020we} and first-principles density functional
theory (DFT) calculations with several different DFT functionals \citep{Hooshmand:2020tu}.}
\label{tab:Tact}
\end{table}

\newpage{}

\newpage{}

\begin{figure}[H]
\centering\leavevmode \includegraphics[width=0.9\textwidth]{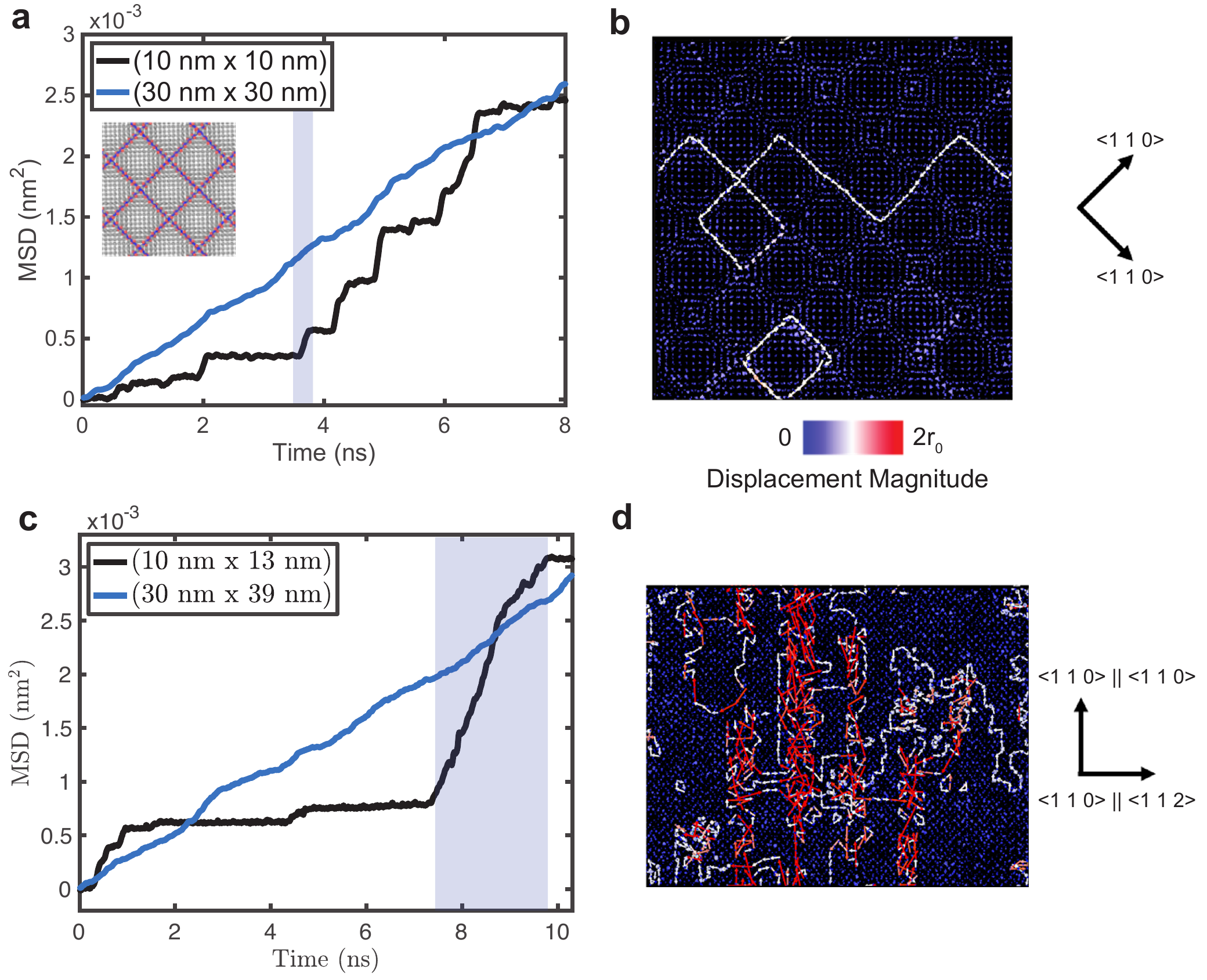}
\caption{Intermittent diffusion behavior in the (a,b) $\Sigma85$ $\{100\}$
low-angle twist Al GB at 500 K and (c,d) \{100\}$||$\{111\} HAIC
Al GB at 700 K. The twist GB is composed of a square network of dislocations
displayed in the top left corner of the MSD plot. Panels (b) and (d)
are atomic displacement maps using the same color coding as in Fig.~\ref{fig:fig3}
and representing the avalanches shown as blue stripes in (a) and (c).
In (b), the GB structure constrains the collective atomic displacements
to the dislocation cores, resulting in string-like and ring-like trajectories.}
\label{fig:fig9}
\end{figure}

\newpage{}
\noindent \begin{center}
\textbf{\LARGE{}SUPPLEMENTARY INFORMATION}{\LARGE\par}
\par\end{center}

\bigskip{}

\textbf{\textcolor{black}{\large{}Point-defect avalanches mediate
grain boundary diffusion}}\textbf{\textcolor{black}{{} }}
\noindent \begin{center}
\textcolor{black}{\large{}Ian Chesser and Yuri Mishin}{\large\par}
\par\end{center}

\newpage{}

\setcounter{figure}{0}

\setcounter{table}{0}

\begin{figure}[H]
\centering\leavevmode \includegraphics[width=0.7\textwidth]{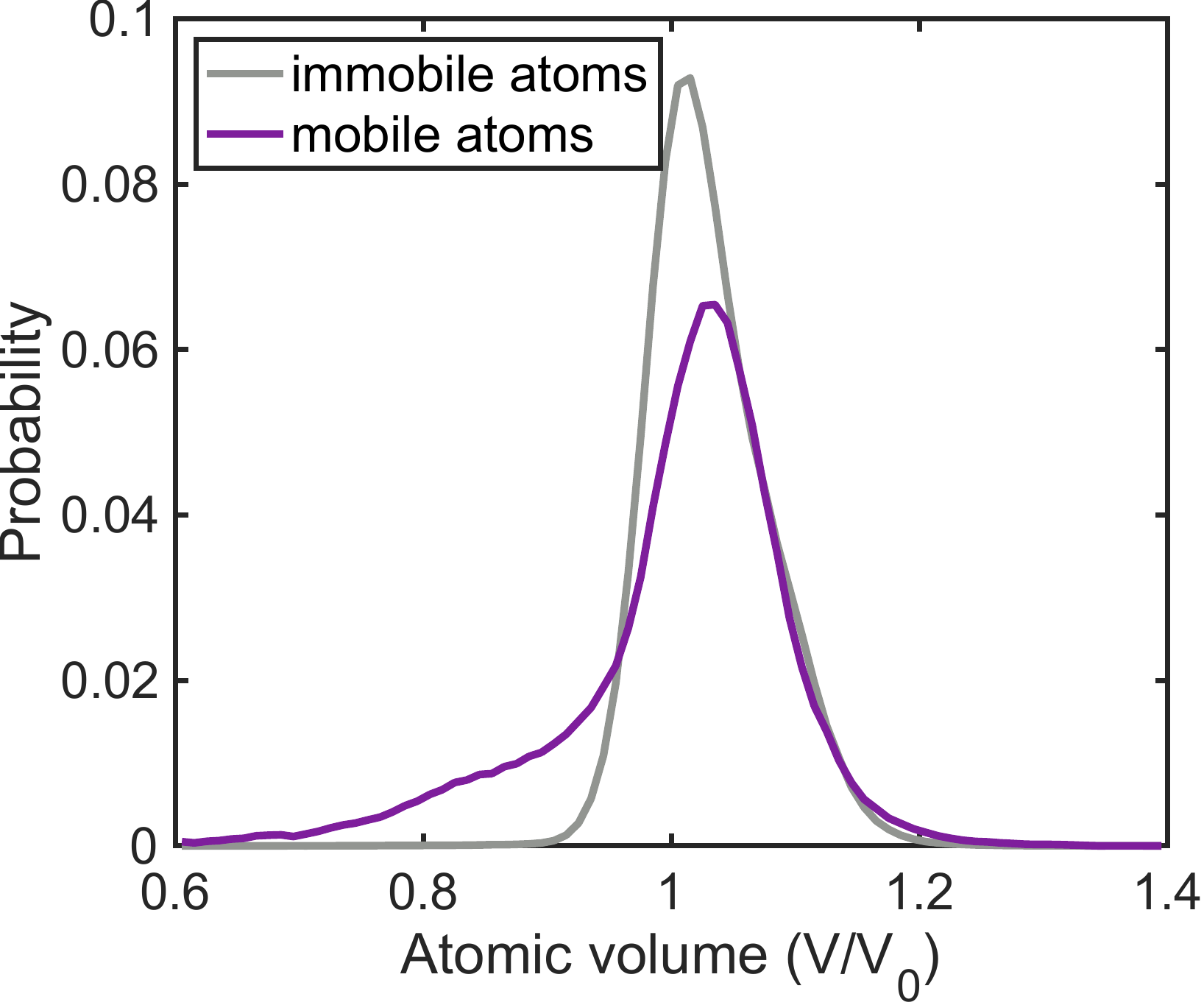}
\caption{Mobile atoms in the $\Sigma3$ Al GB core have a signature in the
free volume distribution relative to immobile atoms. Most atoms in
the GB core are in tension relative to the perfect crystal, but the
mobile atoms exhibit larger compressive and tensile tails in the probability
distribution relative to immobile atoms. This is consistent with string-like
displacements having a compressive head with interstitial character
and a tensile tail with vacancy character. The example distribution
is shown for 710 K but is generic to all temperatures surveyed up
to 750 K.}
\label{fig:fig4}
\end{figure}

\begin{figure}[H]
\centering\leavevmode \includegraphics[width=1\textwidth]{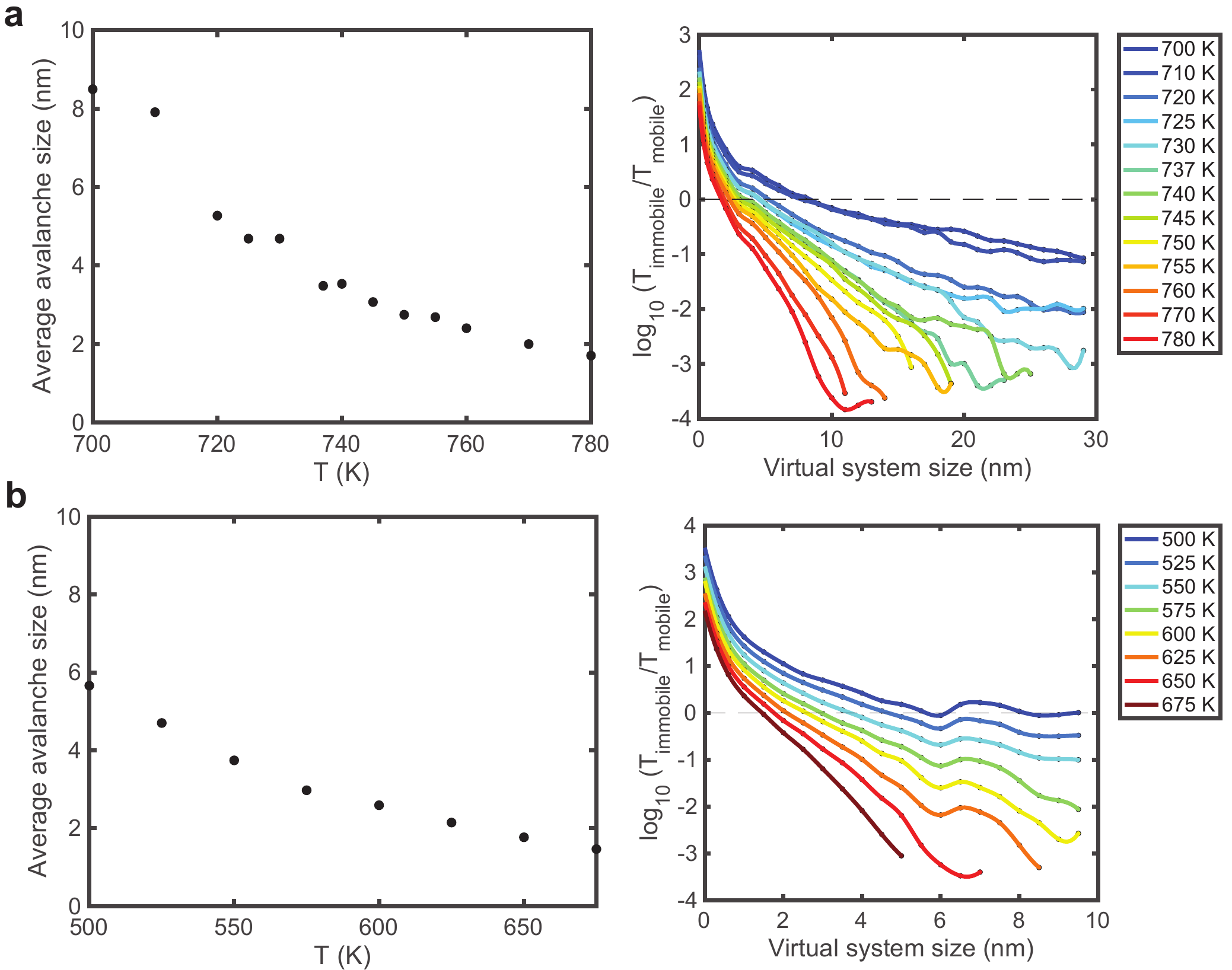}
\caption{Average avalanche size as a function of temperature for the $\Sigma3$
and $\Sigma17$ Al GBs. The avalanche size was extracted from the
virtual system-size analysis by selecting ($l\times l$) GB patches
and measuring the ratio of time spent in locked (immobile) versus
avalanche-like (mobile) states in the space-time trajectories. The
average avalanche size is defined as one corresponding to $t_{\mathrm{immobile}}/t_{\mathrm{mobile}}=1$,
i.e., when the avalanches and locked states occur with equal probability.
The plots demonstrate the decrease in the avalanche size with temperature.}
\label{fig:ava_T}
\end{figure}

\begin{figure}
\noindent \begin{centering}
\includegraphics[width=0.62\textwidth]{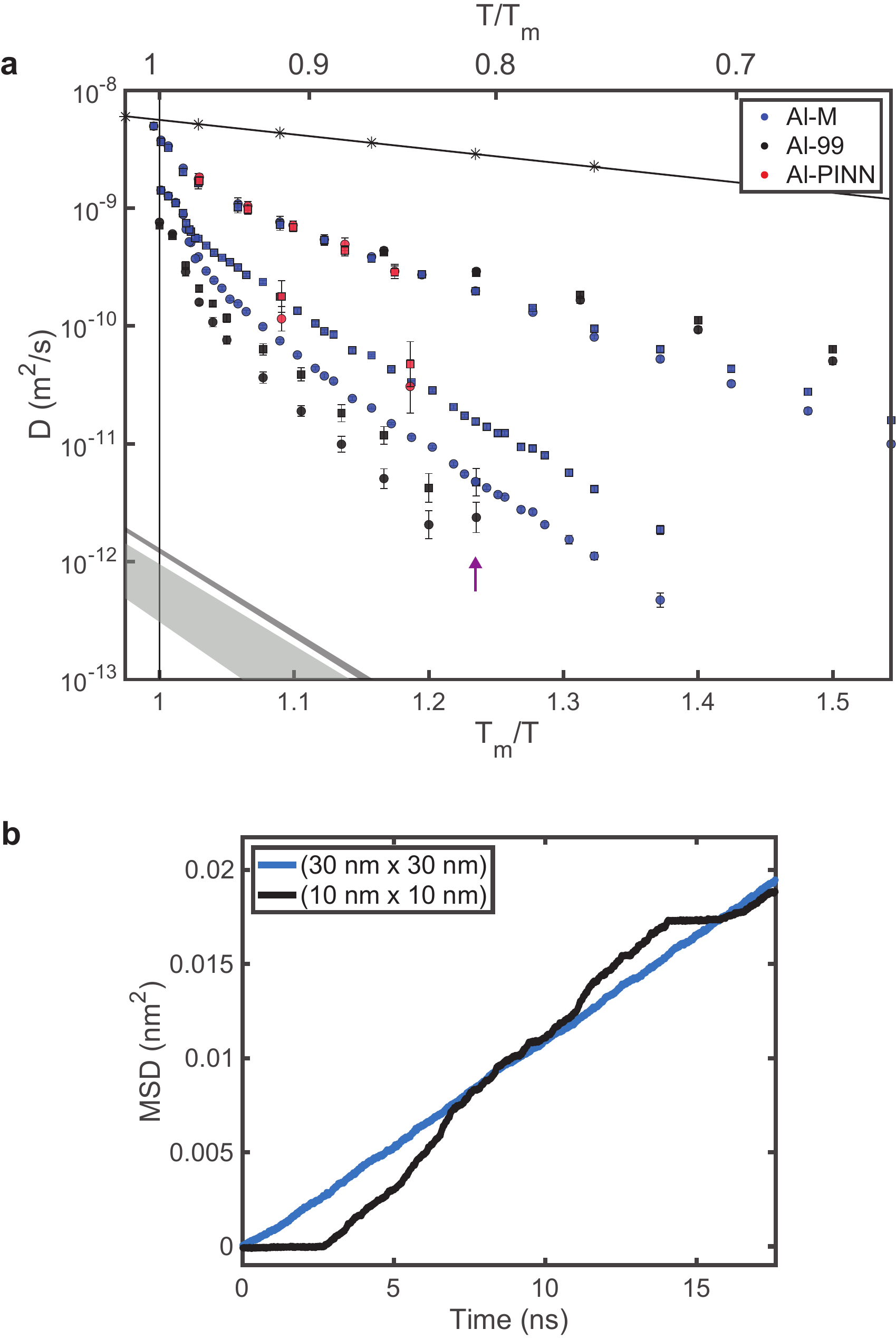}
\par\end{centering}
\caption{(a) Arrhenius diagram of GB diffusion in the $\Sigma3$ and $\Sigma17$
Al GBs computed with three interatomic potentials: Al-M \citep{Almendelev}
(blue), Al-99 \citep{Al99} (black), and Al-PINN \citep{AlPINN2}
(red). The system size is (10 nm $\times$ 10 nm). Due to the slow
computational speed, the Al-PINN calculations only cover a limited
temperature range for the $\Sigma17$ GB and only one temperature
($0.9\ T_{m}$) for the $\Sigma3$ GB. To save computational time,
the GB structures were first equilibrated with the Al-M potential
before switching the atomic interactions to the Al-PINN with an adjustment
of atomic coordinates according to the Al-PINN thermal expansion.
(b) Example of atomic MSD as a function of time at $T=0.82T_{m}$
(point marked by arrow in (a)) computed with the Al-99 potential for
two different GB cross-sections. Note the intermittent behavior for
the smaller cross-section and its disappearance for the larger cross-section.}
\end{figure}

\begin{table}[H]
\centering %
\begin{tabular}{lcccc}
\toprule 
GB description & $x$ & $y$ & $z$ & $\gamma$ (mJ/m$^{2}$)\tabularnewline
\midrule 
\multirow{2}{*}{FCC Ni, $\Sigma3$ $\left\langle 110\right\rangle $ tilt 70.5$^{\circ}$} & $[-1,1,1]$ & $[0,1,-1]$ & $[2,1,1]$ & \multirow{2}{*}{879}\tabularnewline
 & $[1,-1,-1]$ & $[0,-1,1]$ & $[2,1,1]$ & \tabularnewline
 &  &  &  & \tabularnewline
\multirow{2}{*}{BCC Fe, $\Sigma17$ $\left\langle 100\right\rangle $ tilt 61.9$^{\circ}$} & $[5,-3,0]$ & $[0,0,-1]$ & $[3,5,0]$ & \multirow{2}{*}{1096}\tabularnewline
 & $[-5,3,0]$ & $[0,0,1]$ & $[3,5,0]$ & \tabularnewline
 &  &  &  & \tabularnewline
\multirow{1}{*}{} & $[1,1,-2]$ & $[1,-1,0]$ & $[1,1,1]$ & \multirow{1}{*}{}\tabularnewline
\bottomrule
\end{tabular}\caption{Crystallography and 0 K energies ($\gamma$) of the FCC Ni and BCC
Fe GBs studied in this work. The Cartesian $x$, $y$ and $z$ axes
are aligned with edges of the rectangular simulation box. The GB plane
is normal to the $z$-direction. For each GB, the two lines indicate
crystallographic directions parallel to the Cartesian axes in the
upper and lower grains. The GBs were simulated with the EAM interatomic
potentials for FCC Ni \citep{pot_Ni} and BCC Fe \citep{pot_Fe}.
The initial optimized GB structures were taken from the GB structure
databases in \citep{olmsted2009survey2} and \citep{GBE_Fe}.\label{tab:1-1}}
\end{table}

\begin{figure}[H]
\centering\leavevmode \includegraphics[width=0.9\textwidth]{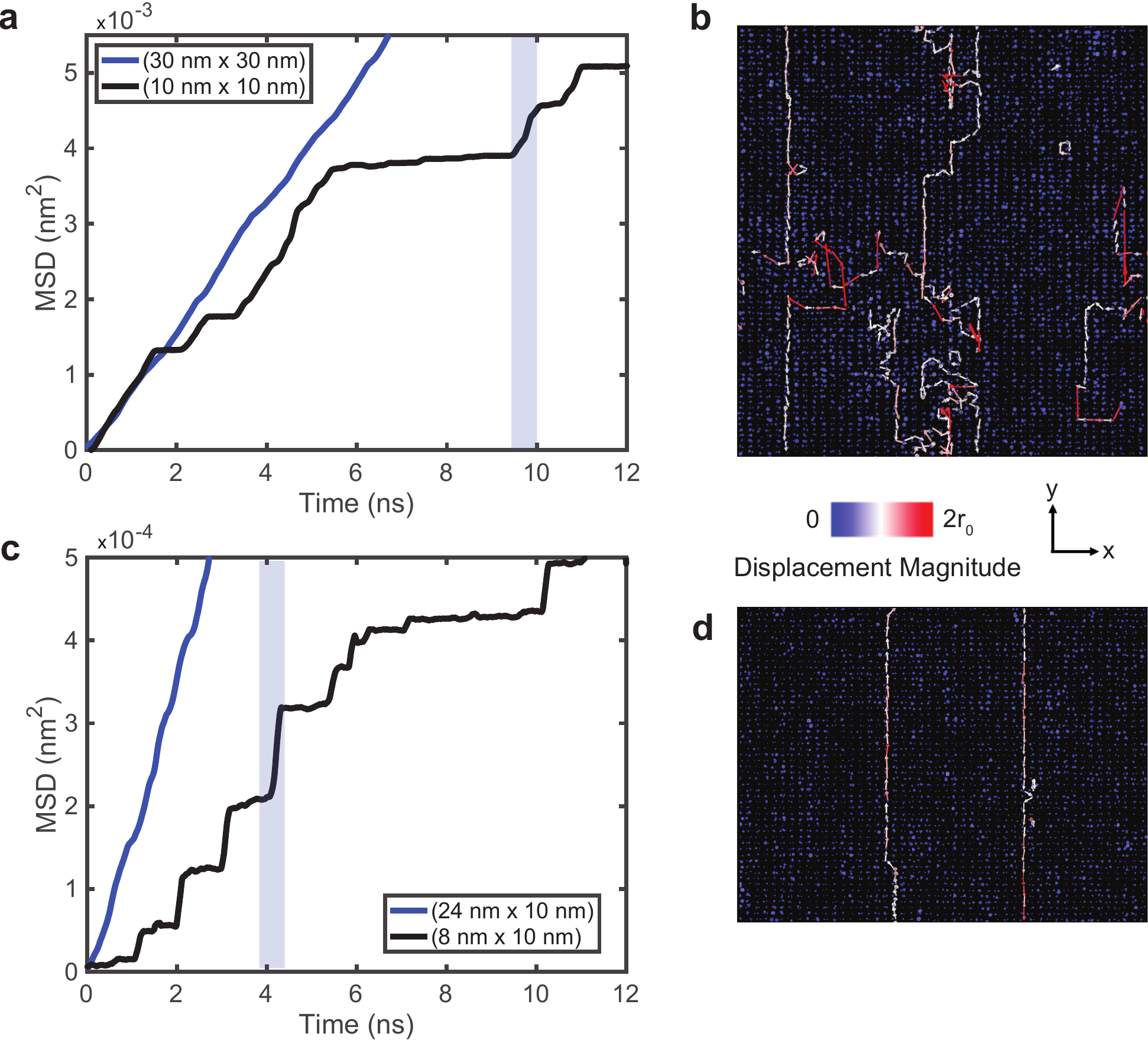}
\caption{Size dependent intermittent diffusion behavior in the (a,b) $\Sigma3$
$\left\langle 110\right\rangle $ tilt 70.5$^{\circ}$ GB in FCC Ni
at 1200 K (0.77 $T_{m}$) and (c,d) the $\Sigma17$ $\left\langle 100\right\rangle $
tilt 61.9$^{\circ}$ in BCC Fe at 1050 K (0.59 $T_{m}$). Panels (b)
and (d) show atomic displacement maps using the same color coding
as in Fig.~2 of the main text and demonstrating the avalanches occurring
in the blue highlighted regions in (a) and (c).}
\label{fig:fig9-1}
\end{figure}

\end{document}